\documentclass[10pt]{article}

\usepackage{amsmath}
\usepackage{amssymb}
\usepackage{graphics}
\usepackage{rotating}
\usepackage{cite}
\usepackage{color}
\usepackage{fancybox}

\def\0{\mbox{\tiny $0$}}
\def\1{\mbox{\tiny $1$}}
\def\2{\mbox{\tiny $2$}}
\def\3{\mbox{\tiny $3$}}
\def\4{\mbox{\tiny $4$}}
\def\5{\mbox{\tiny $5$}}
\def\6{\mbox{\tiny $6$}}
\def\7{\mbox{\tiny $7$}}
\def\8{\mbox{\tiny $8$}}
\def\9{\mbox{\tiny $9$}}

\def\b{\mbox{\tiny $(2)$}}
\def\bd{\mbox{\tiny $(1+1)$}}

\def\d{\mbox{\tiny $(4)$}}

\def\Np{\mbox{\tiny $(N)$}}

\def\T{\mbox{\tiny $T$}}
\def\N{\mbox{\tiny $N$}}
\def\S{\mbox{\tiny $S$}}
\title{\shadowbox{\large \bf
WAVE AND PARTICLE LIMIT FOR MULTIPLE BARRIER TUNNELLING}}

\author{
\small  Stefano De Leo\thanks{Department of Applied Mathematics,
State University of Campinas, Brazil [deleo@ime.unicamp.br] } \,\,
and\, Pietro Rotelli\thanks{Department of Physics, University of
Salento and INFN Lecce, Italy [rotelli@le.infn.it]}}

\date{\small
\fcolorbox{black}{yellow} {\color{red} $\bullet$ {\color{black}{
{\footnotesize {\sc Journal of Physics A} {\bf 44}, 435305-15 (2011)}}} {\color{red}{$\bullet$}} } }

\begin{document}
%
%%%%%%%%%%%%%%%%%%%%%%%%%%%%%%%% PAPER %%%%%%%%%%%%%%%%%%%%%%%%%%%%%%%%%%%%%

\maketitle

\vspace*{-.7cm}

\begin{abstract}
\noindent The particle approach to one-dimensional potential
scattering is applied to non relativistic tunnelling between two,
three and four identical barriers. We demonstrate as expected that
the infinite sum of particle contributions yield the plane wave
results. In particular, the existence of resonance/transparency
for twin tunnelling in the wave limit is immediately obvious. The
known resonances for three and four barriers are also derived. The
transition from the wave limit to the particle limit is exhibit
numerically.
\end{abstract}

%%%%%%%%%%%%%%%%%%%%%%%%%%%%%%%%%%%%%%%%%%%%%%%%%%%%%%%%%%%%%%%%%%%%%%%
%%%%%%%%%%%%%%%%%%%%%%%%%%%%%%%%%%%%%%%%%%%%%%%%%%%%%%%%%%%%%%%%%%%%%%%

%%%%%%%%%%%%%%%%%%%%%%%%%%%%%%%%%%%%%%%%%%%%%%%%%%%%%%%%%%%%%%%%%%%%%%%
%%%%%%%%%%%%%%%%%%%%%%%%%%%%%%%%%%%%%%%%%%%%%%%%%%%%%%%%%%%%%%%%%%%%%%%

% Graphene electronic transport, 72.80.Vp
% Graphene films, 68.65.Pq

% Warning: No PACS code given

%02.10.Hh Rings and algebras
%02.10.Ud Linear algebra
%02.10.Yn Matrix theory

%02.30.Hq Ordinary differential equations
%02.30.Jr Partial differential equations
%02.30.Tb Operator theory

%03.65.-w Quantum mechanics
%03.65.Ca Formalism
%03.65.Ta Foundations of quantum mechanics;
%03.65.Xp Tunnelling, traversal time, quantum Zeno dynamics

%12.15.F Quarks and lepton masses and mixing
%14.60.Pq Neutrino mass and mixing

%\offprints{~Stefano De Leo.}

%%%%%%%%%%%%%%%%%%%%%%%%%%%%%%%%%%%%%%%%%%%%%%%%%%%%%%%%%%%%%%%%%%%%%%%%%%%%

%
%%%%%%%%%%%%%%%%%%%%%%%%%%%%%%%% PAPER %%%%%%%%%%%%%%%%%%%%%%%%%%%%%%%%%%%%%

%%%%%%%%%%%%%%%%%%%%%%%%%%%%%%%%%%%%%%%%%%%%%%%%%%%%%%%%%%%%%%%%%%%%%%%
%%%%%%%%%%%%%%%%%%%%%%%%%%%%%%  SECTION   %%%%%%%%%%%%%%%%%%%%%%%%%%%%%
%%%%%%%%%%%%%%%%%%%%%%%%%%%%%%%%%%%%%%%%%%%%%%%%%%%%%%%%%%%%%%%%%%%%%%

\section*{\normalsize I. INTRODUCTION.}

The particle approach to piece-wise potential scattering is
characterized by considering the reflection and transmission
amplitudes, successively, at each potential discontinuity. this
method, applied also in optics\cite{BORN}, can be used as an
alternative to plane wave continuity equations for the derivation
of the transmission and reflection amplitudes and find application
in approximate solutions to one-dimensional potential problems
\cite{Fermor,Anderson}. Several interesting papers have analyzed
tunneling with this method\cite{McVoy,Esposito,Cardone}. However,
in certain situations, this approach is the natural physical
choice in the ``particle limit" in which the incoming wave packets
are small compared to the potentials extension. However, it must
be emphasized that the approach itself does not require the use of
wave packets. It predicts multiple reflections and even for a
single incoming wave it generally results in infinite reflected
and transmitted waves. An example of this approach is the
diffusion above a single potential barrier, from which
one significant consequence, otherwise mysterious, is the
step limit for large barrier\cite{DifS}. The first barrier reflection
coefficient reproduces the step result.

 The alternative and standard approach to potential scattering is
 with a single  wave analysis\cite{Cohen}. We refer to this as the ``wave limit".
 It is characterized by continuity equations often best described
 by the use of matrices and yields a {\em single} reflected and
 transmitted amplitude. Due to the fact that the algebraic sum of the infinite
 contributions in the particle approach yields the wave result,
 the two methods are mathematically equivalent.
 However, they are not equivalent in practice because
 a sum of
 calculated terms, if finite, is an unambiguous process, but the decomposition
 of an expression as an infinite sum, on the other hand, is
 highly ambiguous, even if only one decomposition can lead to probability
 conservation and to the correct exit times of the separate wave packets.
In the wave limit, probabilities are calculated by first summing
the particle terms and then squaring the modulus. In the particle
limit, the probabilities are calculated by first squaring the
modulus of each term and then summing.

Resonance phenomena, within the Schr\"odinger equation, are well
known for a single barrier when considering plane waves with
energy above the potential value, i.e.  $E>V_{\0}$. It results in
unit transmission probabilities. For a given plane wave energy
there are unlimited such resonances as the barrier length
increases. More realistically, for an incoming  wave packet it can
be shown that resonance occurs only if the barrier length is much
smaller than the wave packet dimensions\cite{DifD}. Resonance is a
property of the wave nature of particles while, in the other limit
in which the wave packet is small compared to the barrier length,
the transmission amplitude {\em breaks up} into infinite
transmitted (and consequently reflected) wave packets, the so
called ``particle limit". There is no particle limit for
tunnelling, $E<V_{\0}$,
 through a single barrier\cite{HE,Rec,Win}. However, for more than one
barrier there exists another coherence effect, tunnelling
resonances\cite{Hau,Jap,DB,Aha,TunS,Res}. For particular values of
the momentum and of the inter-barrier distance,  identical
barriers become {\em transparent}, i.e. the tunnelling
transmission probability equals unity.

In the next section, we recall the single barrier tunnelling
results and define the amplitudes relevant to the particle
approach. We then consider the case of twin barriers. We
illustrate our procedure for calculating the individual particle
contributions and then sum them to obtain the wave limit result.
This sum can also be viewed as a compact expression for the
particle series. The resonance condition will be obvious  in this
expression. We shall also  prove probability conservation in both
the wave and particle limit. Conservation of probability must also
be valid for all intermediate (partial wave packet overlap) cases,
however no simple analytic calculation of these probabilities is
known. In section III, we describe some numerical calculations in
which an incoming Gaussian wave packet is used. The existence,
albeit degraded, of resonances is indicted by oscillations as a
function of inter-barrier distance $d$. For large $d$, the
particle picture sets in and the oscillatory behavior is {\em
damped out}. In section IV, we consider the three identical
barrier case and derive the, non obvious, resonance conditions. To
proceed to higher numbers of barriers it is more practical  to
change the method of calculation. In section V, we derive  a
matrix expression for the wave limit and then describe how this
can be rewritten in a form from which the individual particle
series can be extracted. This is applied, as an example, to the
case of four identical barriers and the first transmitted particle
terms confirm the direct particle procedure. We draw our
conclusions in section VI.

\section*{\normalsize II. TWIN BARRIER TUNNELLING}

To perform our calculation, we need the reflection and
transmission amplitudes for a single barrier.  The single barrier
reflection amplitude depends not only upon the barrier height and length but also on
the barrier  position and consequently upon whether the barrier is
encountered coming from the left or from the right. Nevertheless,
the explicit rules are quite simple.

Let $R$ be the  reflection amplitude for a plane wave of unit
amplitude  with energy $E<V_{\0}$ impinging on a barrier of length
$L$ from the left at $x=0$, see Fig.\,1-a. Let $T$ be the
transmission amplitude  for the same wave. Notice that the
reflection and transmission probabilities, also known as the
reflection and transmission coefficients, are consequently
$|R|^{^{2}}$ and $|T|^{^{2}}$ respectively. Now, a standard plane
wave calculation gives the results
\begin{eqnarray}
R & = & - \,i \, \frac{k^{\2}+\rho^{\2}}{2k\rho}\,\cos \phi\,\tanh
(\rho L) \,\exp(i \phi) = - i\,|R| \,\exp(i \phi)\,\,,
  \\
T & = & \frac{\cos \phi}{\cosh(\rho L)}\, \exp[i(\phi-kL)] = |T|\,
\exp[i(\phi-kL)]\,\,, \label{teq}
\end{eqnarray}
with \[ k=\sqrt{2mE}/\hbar\,\,\,,\,\,\,\,\,
\rho=\sqrt{2m(V_{\0}-E)}/\hbar\,\,\,\,\, \mbox{and} \,\,\,\,\,\tan
\phi = ( k^{\2}-\rho^{\2})\tanh (\rho L) /\, 2k \rho\,\,,\] where,
by convention,
\[ -\,\pi/2<\phi<\pi/2\,\,.\]
Conservation of probability results from the fact that
$|R|^{^{2}}+|T|^{^{2}}=1$. Now, shift the front of the barrier
from the origin to the position $x=a$. The reflection amplitude
acquires an additional phase, becoming
\begin{equation}
R\,\exp(2\,ik\,a)\,\,.
\end{equation}
 If the wave impinges the barrier from the right (momentum $-k$), we
obtain instead
\begin{equation}
R\,\exp(-2\,ik\,b)\,\,,
\end{equation}
where $b=a+L$. In both cases the transmission amplitude remains
{\em unchanged}. It does not depend upon the position of the
barrier, but only upon its width ($L$) and height ($V_{\0}$),
apart, of course, upon the value of the energy $E$.

Now consider the twin barrier problem,  the shape of the potential
is drawn in Fig.\,1-b. The particle approach stems from
considering the hypothetical case of wave packets such that their
spatial dimensions are small compared to $d$.

Consider an incoming wave packet from the left. When it reaches
the first barrier at $x=0$, it will produce a first reflected wave
in the region $x<0$  and a first transmitted wave into the the
inter-barrier separation. For {\em each} momentum component $k$,
these are given by the
\[ R \,\,\,\,\,\mbox{and}\,\,\,\,\, T\]
of the single barrier given above, with consequent
conservation of probability. Now the wave packet in the
inter-barrier region will travel to the second barrier and then be
partially reflected by and partially transmitted through the
second barrier. These amplitudes will be, respectively,
 \[ T\times R\,e^{^{2ik(L+d)}} \,\,\,\,\,\mbox{and}\,\,\,\,\, T\times T\,\,.\]
The latter amplitude is the first contribution to the total
transmission amplitude. Again probability is clearly conserved at
this step. The reflected wave from the second barrier now
travelling to the left impinges upon the first barrier at $x=L$
and, after its transmission through the first barrier, creates the
second contribution to the total reflection amplitude,
\[ TR\,e^{^{2ik(L+d)}}\times T\,\, .\]
The reflected amplitude (at $x=L$)  which moves back from the
first to second barrier is
\[
TR\,e^{^{2ik(L+d)}}\times R\,e^{^{-2ikL}}\,\,.
\]
Again probability is conserved. Consequently, the second
contribution to the total transmission amplitude is then given by
\[
TR^{^{2}}\,e^{^{2ikd}}\times T\,\,.
\]
The procedure repeats continuously and the sum of these
contributions gives
\begin{eqnarray}
R_{\S}^{^{(2)}} & = & R+
R\,T^{^{2}}e^{^{2ik(L+d)}}\,\sum_{n=0}^{\infty}
\left(R^{^{2}}\,e^{^{2ikd}}\right)^{^{n}} =  R
+R\,T^{^{2}}e^{^{2ik(L+d)}}\, \mbox{\Large $/$}
  \left(1-R^{^{2}}\,e^{^{2ikd}}\right) \label{refl}\,\,,\\
T_{\S}^{^{(2)}} & = & T^{^{2}} \sum_{n=0}^{\infty}
\left(R^{^{2}}\,e^{^{2ikd}}\right)^{^{n}} = T^{^{2}}\,
\mbox{\Large $/$}
  \left(1-R^{^{2}}\,e^{^{2ikd}}\right)\,\,, \label{tra}
\end{eqnarray}
where the bracketed upper script $(2)$ on the l.h.s. indicates the
number of identical barriers and the subscript S refers to an infinite sum.
The second equality in each of the
above equations is legitimate because $|R^{^{2}}\,e^{^{2ikd}}|<1$.

For an incoming wave packet which is small in spatial dimensions
compared to the inter-barrier potential distance, the individual
contributions used in the sums above should not be added because
they are incoherent. Each will correspond to a separate outgoing
(transmitted or reflected) wave packet,
\begin{eqnarray}
\mathcal{R}_{\mbox{\tiny particle}}^{^{{(2)}}} & = &
|R|^{^{2}}+|R|^{^{2}}|T|^{^{4}} \sum_{n=0}^{\infty} |R|^{^{4 n}} =
2\,|R|^{^{2}}\mbox{\Large $/$}
  \left(1+|R|^{^{2}}\right) \label{prefl}\,\,,\\
\mathcal{T}_{\mbox{\tiny particle}}^{^{{(2)}}} & = & |T|^{^{4}}
\sum_{n=0}^{\infty} |R|^{^{4 n}} = |T|^{^{2}}\mbox{\Large $/$}
  \left(1+|R|^{^{2}}\right) \label{ptra}\,\, .
\end{eqnarray}
This is what we call the {\em particle limit}. Since probability
is conserved at each step of the above calculation, probability is
conserved overall,
\[
\mathcal{R}_{\mbox{\tiny  particle}}^{^{(2)}}+\mathcal{T}_{\mbox{\tiny
particle}}^{^{{(2) }}} =  1\, \, ,
\]
but the transmission and reflection probabilities in the {\em
particle limit} are not those of the {\em wave limit} when each
individual particle contribution is completely
coherent\cite{DB,TunS,Res}. In this case,
\begin{eqnarray}
\mathcal{R}_{\mbox{\tiny wave}}^{^{{(2) }}} & = & 2\,|R|^{^{2}}\,
[1+\cos(2\alpha)]\,\mbox{\Large $/$}
  \left[1+|R|^{^{4}}+2\,|R|^{^{2}}\cos(2\alpha)\right]
 \label{preflw}\,\,,\label{prw2}\\
\mathcal{T}_{\mbox{\tiny wave}}^{^{{(2) }}} & = & |T|^{^{4}}\,
\mbox{\Large $/$}
  \left[1+|R|^{^{4}}+2\,|R|^{^{2}}\cos(2\alpha)\right] \label{ptraw}\,\,
  ,
\end{eqnarray}
with
\[ \alpha = \phi + kd \,\,,\]
and, as expected,
\[
\mathcal{R}_{\mbox{\tiny wave}}^{^{{\b }}}+\mathcal{T}_{\mbox{\tiny wave}}^{^{{\b
}}}=1\,\, .\]
 This result must not be misinterpreted as proof of
a single outgoing transmission/reflection amplitude. The
transition between the wave and the particle limit will be
discussed in more detail in the next section.

Now let us return to the transmission amplitude given in
Eq.(\ref{tra}), seen in the wave limit,
\begin{equation}
 T_{\S}^{^{(2)}} = T^{^{2}}\, \mbox{\Large $/$} \,D\,\,,
\end{equation}
with
\[ D = 1- R^{^{2}}e^{^{2ikd}}=1+|R|^{^{2}}e^{^{2i\alpha}}\,\,. \]
From this form, one immediately derives the condition for
resonance tunnelling for which the barriers becomes
``transparent". The {\em maximum} of the above expression occurs
when the modulus of the denominator is a minimum, i.e. when
\begin{equation}
\cos(2\alpha)=-1\,\,\,\Leftrightarrow\,\,\,
\cos(\alpha)=0\,\,.\end{equation}
 For these values of $\alpha$
\begin{equation}
 T_{\S}^{^{(2)}}\left[\cos \alpha=0\right] =  T^{^{2}} \mbox{\Large
 $/$}\,
 |T|^{^{2}}\,\,\,\,\,\Rightarrow \,\,\,\,\,
 \mathcal{T}_{\mbox{\tiny wave}}^{^{{(2)}}}\left[\cos \alpha=0\right] =
1\,\,.
\end{equation}
Observe that the transmission probability in the particle limit,
Eq.(\ref{ptra}), can {\em never} be unity, since $|T|<1$. We may
also determine the {\em minimum} value of $P_{\mbox{\tiny
T,wave}}^{^{{\b}}}$ which occurs when
\begin{equation}
\cos(2\alpha)=+\,1\,\,\,\Leftrightarrow\,\,\, \cos(\alpha)=\pm\,
1\,\,.\end{equation}
For these values of $\alpha$
\begin{equation}
 T_{\S}^{^{(2)}}\left[\cos \alpha=\pm 1\right] =  T^{^{2}} \mbox{\Large
 $/$}\,\left(1+
 |R|^{^{2}}\right)\,\,\,\,\,\Rightarrow \,\,\,\,\,
 \mathcal{T}_{\mbox{\tiny wave}}^{^{(2)}}\left[\cos \alpha=\pm 1\right]
=1\, \mbox{\Large
 $/$}\,
\left(1+
 2\,|R|^{^{2}}/|T|^{^{2}}\right)^{\2}\,\,.
\end{equation}

\section*{\normalsize III. TRANSITION BETWEEN WAVE AND PARTICLE LIMIT}

In this section, we intend to study numerically  the behavior of a
particular incident wave packet impinging on a double identical
barrier potential (Fig.\,1-b) from the left. The results will
graphically exhibit the transition from the wave to particle
limits and also verify the validity of some of the expressions
derived in the previous section. For the following numerical
calculation, we shall express all quantities in terms of the
barrier height $V_{\0}$ and/or $mV_{\0}$. For our purposes these
values need not to be explicitly fixed. We shall consider below a set
of barrier widths, the incoming wave packet's mean momentum
($k_{\0}$)/energy ($E_{\0}$) and a set of its momentum spread. The
inter-barrier distance will be left as a continuous variable in
our calculations. The incident wave packet is obtained by
superposing the planes waves
\[ \exp\left[ i \left(k\, x - \frac{E\,t}{\hbar}\right)\right]\]
with the coefficient
\[
\frac{\sqrt{a}}{(2\pi)^{^{3/4}}}\,
\exp\left[-\,\frac{\,a^{\2}}{4}\left(k-k_{\0}\right)^{\2}\right]\,\,,
\]
which correspond to a Gaussian function centered at
$k_{\0}=\sqrt{2mE_{\0}}/\hbar$ multiplied by  a numerical factor
which normalizes the wave function\cite{Cohen}. The transmitted
wave packet can be written as
\begin{equation}
\label{wpa} \Psi_{tra}(x,t)=
\frac{\sqrt{a}}{(2\pi)^{^{3/4}}}\,\int_{_{0}}^{^w}\mbox{d}k\,\,
T_{\S}^{^{(2)}}\,
\exp\left[-\,\frac{\,a^{\2}}{4}\left(k-k_{\0}\right)^{\2}+
i\left(k\,x-\frac{E\,t}{\hbar}\right)\right]\,\,,
\end{equation}
where $w=\sqrt{2mV_{0}}/\hbar\,$.
%In order to simplify our expressions, we introduce the following
%adimensional quantities
%\[
%\epsilon=\frac{\hbar \,k}{\sqrt{2mV_{\0}}}\,\,\,,\,\,\,\,\,
%\epsilon_{\0}=\frac{\hbar \,k_{\0}}{\sqrt{2mV_{\0}}}
%\,\,\,,\,\,\,\,\, \left(\xi,\sigma,\lambda,\delta\right) =
%\frac{\sqrt{2mV_{\0}}}{\hbar}\left(x,a,L,d\right)\,\,\,\,\,\mbox{and}\,\,\,\,\,
%\tau=\frac{V_{\0}t}{\hbar}\,\,.
%\]
%By using this new notation, the wave packet amplitude (\ref{wpa})
%becomes
%\begin{eqnarray}
%\Psi_{tra}(x,t) & = &
%\frac{\sqrt{a}}{(2\pi)^{^{3/4}}}\,\frac{\sqrt{2mV_{\0}}}{\hbar}\,\int_{_{0}}^{^{1}}
%\mbox{d}\epsilon \,\, T_{\S}^{^{(2)}}[\epsilon,\lambda,\delta\,]\,
%\exp\left[-\,\frac{\,\sigma^{\2}}{4}\left(\epsilon-\epsilon_{\0}\right)^{\2}+
%i\left(\epsilon\,\xi-\epsilon^{\2}\,\tau\right)\right] \nonumber
%\\
% & = & \frac{\sqrt{a}}{(2\pi)^{^{3/4}}}\,\frac{\sqrt{2mV_{\0}}}{\hbar}\,\int_{_{0}}^{^{1}}
%\mbox{d}\epsilon
%\,\,g\left[\epsilon,\lambda,\delta,\epsilon_{\0},\sigma,\xi,\tau\right]\,\,,
%\end{eqnarray}
%with
%\[T_{\S}^{^{(2)}}[\epsilon,\lambda,\delta\,]=T^{^{2}}[\epsilon,\lambda]\,
%\mbox{\Large $/$}
%  \left(1-R^{^{2}}[\epsilon,\lambda]\,e^{^{2i\epsilon\delta}}\right)\,\,.
%\]
Consequently, the probability of transmission is given by
\begin{equation} \int_{_{2L+d}}^{^{+
\,\infty}}\hspace*{-.3cm}\mbox{d}x\,\,\left|
\Psi_{tra}(x,t)\right|^{^{2}}\,\,.
\end{equation}
%where we maintain for
%convenience the integral over $x$ instead of writing it as an
%integral over $\xi$. Explicitly this reads,
%\begin{equation}
%\frac{a}{(2\pi)^{^{3/2}}}\,\frac{2mV_{\0}}{\hbar^{^{2}}}\,
% \int_{_{2L+d}}^{^{+ \,\infty}}\hspace*{-.3cm}\mbox{d}x\,\int_{_{0}}^{^{1}}
%\mbox{d}\epsilon \,\int_{_{0}}^{^{1}} \mbox{d}\epsilon' \,\,
%g^*\left[\epsilon',\lambda,\delta,\epsilon_{\0},\sigma,\xi,\tau\right]
%g\left[\epsilon,\lambda,\delta,\epsilon_{\0},\sigma,\xi,\tau\right]
%\end{equation}
Now we are interested in the {\em total} transmission probability.
This is formally the limit of the above for $t \to \infty$.
However, there is an alternative procedure to obtain the same
result. We send the lower $x$ limit to $- \infty$. This picks up
all the future (at time $t$) probability contributions, i.e. {\em
phantom} wave packets. An example of this phenomena is seen
 in numerical calculations with, say, the step potential
 for times before total ($E<V_{\0}$)  wave packet reflection. If one
 plots the reflected wave amplitude, it does not appear yet in the free
 potential region where destructive interference occurs but a
 wave does appear for $x$ values in the {\em forbidden} region.
Of course, the reflection coefficient does not apply there, and
this traveling wave packet is only virtual or phantom. However,
it eventually emerges as a physical wave packet in the free region
at the appropriate reflection time. The above extension of the
lower $x$ integration limit is useful for Gaussian convolution
 because it yields a Dirac delta
function. This allows us to immediately perform one of the
momentum integrals. It in turn eliminates the time dependence as
it must, and yields the following expression for the total
transmission probability,
\begin{eqnarray}
\mathcal{P}_{\T}^{^{(2)}} & = &\frac{\sqrt{2mV_{\0}}\,a/\hbar}{\sqrt{2\pi}}\,\int_{_{0}}^{^{1}}
\mbox{d}\left(\frac{\hbar\,k}{\sqrt{2mV_{\0}}}\right)\,
\,\,\left|\,\,T_{\S}^{^{(2)}}\right|^{^{\,2}}\,
\exp\left[-\,\frac{(\sqrt{2mV_{\0}}\,a/\hbar)^{^{2}}}{2}\,
\left(\frac{\hbar\,k-\hbar\,k_{\0}}{\sqrt{2mV_{\0}}}\right)^{\2}\right] \nonumber \\
 & = & \frac{\sqrt{2mV_{\0}}\,a/\hbar}{\sqrt{2\pi}}\,\int_{_{0}}^{^{1}}
\mbox{d}\sqrt{\frac{E}{V_{\0}}}\,
\,\,\left|\,\,T_{\S}^{^{(2)}}\right|^{^{\,2}}\,
\exp\left[-\,\frac{(\sqrt{2mV_{\0}}\,a/\hbar)^{^{2}}}{2}\,
\left( \sqrt{\frac{E}{V_{\0}}} -   \sqrt{\frac{E_{0}}{V_{\0}}} \, \right)^{\2}\right]\,\,.
\end{eqnarray}
From the previous equation, observing that $T_{\S}^{^{(2)}}$ is a function of
$E/V_{\0}$, $\sqrt{2mV_{\0}}\,d/\hbar$ and  $\sqrt{2mV_{\0}}\,L/\hbar$,
we immediately see that $\mathcal{P}^{^{(2)}}_{\T}$ is completely determined once fixed the following four adimensional quantities,
\[ \frac{E_{\0}}{V_{\0}}\,\,,\,\,\,\,\,
\frac{\sqrt{2mV_{\0}}\,a}{\hbar}
\,\,,\,\,\,\,\,
\frac{\sqrt{2mV_{\0}}\,d}{\hbar}
\,\,\,\,\,\,\,\mbox{and}\,\,\,\,\,\,\,
\frac{\sqrt{2mV_{\0}}\,L}{\hbar}\,\,.
\]
$\mathcal{P}^{^{(2)}}_{\T}$ is calculated numerically and plotted in
Fig.\,2, where we have chosen the ratio $E_{\0}/V_{\0}$ equals to $1/2$.
In the upper part of the figure, Fig.\,2-a, we set the barrier
width,
\[ \sqrt{2mV_{\0}}\,L/\hbar =1\,\,,\]
 and display the total transmission
probability versus the inter-barrier distance,
$\sqrt{2mV_{\0}}\,d/\hbar$, for various wave packet widths. The wave limit is obtained when each individual contribution to $\left|\,\,T_{\S}^{^{(2)}}\right|$ is completely coherent. This happens when the inter-barrier potential distance ($d$) is small if compared to the wave packet spatial dimension ($a$),
\[ \mathcal{P}_{\T}^{^{(2)}}[\,a\gg d\,] \,\,\rightarrow\,\,  \left[\, \mathcal{T}_{\mbox{\tiny wave}}^{^{{\b
}}}\right]_{\0}\,\,,  \]
where the subscript $0$ indicates that $\mathcal{T}_{\mbox{\tiny wave}}^{^{{\b
}}}$ at $E=E_{0}$. For
plane waves, we would have a periodic resonance structure with
maxima at
\[ \sqrt{2mV_{\0}}\,d/\hbar =(2n+1)\,\pi\,/\,\sqrt{2}\,\,.\]
As shown in Fig.2-a, the only close approximation to a resonance/transparency occurs
for the first maximum and for the largest spreads of wave packet
plotted. As $\sqrt{2mV_{\0}}\,d/\hbar$ increases in our plot, for
each choice of wave packet width the probability tends to a
constant value. This constant value is equal to $0.46$ and it is in
agreement with the particle limit given in Eq.(\ref{ptra}). This is due to
fact that for an incoming wave packet which is small in spatial dimensions ($a$) compared
to the inter-barrier potential distance ($d$) the contributions to $\left|\,\,T_{\S}^{^{(2)}}\right|$ are incoherent,
\[ \mathcal{P}_{\T}^{^{(2)}}[\,a\ll d\,] \,\,\rightarrow\,\,  \left[\mathcal{T}_{\mbox{\tiny particle}}^{^{{\b}}}\right]_{\0}\,\,.  \]
In Fig.\,2-b, we set the wave packet width,
\[ \sqrt{2mV_{\0}}\,a/\hbar =30\,\,,\]
 and display the total transmission
probability versus the inter-barrier distance,
$\sqrt{2mV_{\0}}\,d/\hbar$, for various barrier widths. Again the
gradual transition from wave to particle limit is exhibited by the
damping of the oscillations. Each particle limit is different
because the values of $R$ and $T$ depend upon the barrier width.
Also it is to be noted that the maxima and minima of each curve
occur at the same values of $\sqrt{2mV_{\0}}\,d/\hbar$. This is
{\em not} true in general. It happens here because of our choice
of $E_{\0}=V_{\0}/2$. This particular ratio results in $\phi$=0.
In general $\phi$ is dependent upon the barrier width and this
would separate somewhat the maxima/minima.

The curves, plotted in Fig.\,2, clearly show the transition from
the wave to the particle limit. However, another even more direct
way to show the particle limit is to display the probability
density as a function of $x$ for the transmitted wave. In Fig.\,3,
this is shown for a case in which multiple wave packets exist and
 for a time at which two have emerged. These are numerical
calculations and we take the opportunity to compare the position
of these first two maxima with that predicted by the stationary
phase method, SPM, approximation\cite{Spm} applied to the particle
expression of Eq.(\ref{tra}),
\[ T_{\S}^{^{(2)}}  =  T^{^{2}}  +  T^{^{2}} R^{^{2}}\,e^{^{2ikd}} + ...\,\,.\]
This method gives us the times of exit of the maxima and, knowing
the group velocity  $\hbar k_{\0} /m$, we can calculate their
later positions. The phase of the first transmitted wave packet is
extracted from
\[ T^{^{2}}\, e^{^{i(kx-Et/\hbar)}}\,\,.\]
The SPM then gives the position of the first maximum at time $t$,
\[ x_{\1} = 2\,L-2\left.
\frac{\partial \phi}{\partial k}\right|_{\0} + \frac{\hbar
k_{\0}}{m}\,t \,\,,\] with the derivative calculated at the
maximum of the Gaussian distribution, $k_{\0}$. In this step we
have neglected the shift in this maximum produced by the
transmission amplitude. The derivative term is proportional to
twice the transition time through a single barrier. It is thus
proportional to the time needed to tunnel through the twin
barriers. Note that this contribution exists even if, as in our
case, $\phi=0$. The SPM transition time for a single barrier is
surprising because for very large $L$ it becomes independent of
the barrier width. This is the Hartman effect\cite{HE}, which has
become of renewed interest in recent years\cite{Rec,Win}.

The phase of the second transmitted wave packet is obtained from
\[ T^{^{2}}\, e^{^{i(kx-Et/\hbar)}}\,\times\, \,R^{^{2}}\,e^{^{2ikd}}\,\,,\]
whence
\[ x_{\2} = x_{\1} -2\left. \frac{\partial \phi}{\partial k}\right|_{\0} - 2\,d\,\,.\]
The derivative terms here are related to the reflection {\em
delay} times and, as is well known, are seen to be equal to the
tunneling times. These SPM position of the maxima, for the case
plotted in Fig.\,3, are indicated by the vertical lines. Agreement
with the numerical calculation is excellent.

However, notice that if the SPM is applied to the summed (wave
limit) expression it will yield a single maximum which does {\em
not} coincide with any of the physical particle maxima. The use of
the SPM must be augmented with, at the very least, a knowledge of
the number of maxima involved.

\section*{\normalsize IV. THREE  BARRIER ANALYSIS}

The three identical barrier problem, Fig. 1-(c), is somewhat more
complicated than the twin barrier case, particularly in the
extraction of the conditions for resonance/transparency. We have
now two identical inter-barrier regions and hence different
processes or paths can contribute to coherent outgoing
``particles", i.e. with the same exit times. These terms  must be
summed before squaring the amplitudes. For brevity, we derive only
the transmission amplitudes.

We proceed as follows. The leading transmission term is that
without any internal reflection, i.e. $T^{^{3}}$. After including
all internal reflections, we find
\begin{equation}
\label{t3} T^{^{3}}\left\{1+2\, R^{^{2}}
 e^{^{2ikd}}+ 3\, R^{^{4}}
 e^{^{4ikd}} + ...\right\}\,\,.
\end{equation}
The factor two  before the $T^{^{3}}R^{^{2}}
 e^{^{2ikd}}$ term allows for a double reflection, $R^{^{2}}$, in each of the inter-barrier
 regions.
 The coefficient (three) in the third term allows for a four-fold
 reflection, $R^{^{4}}$, in each of the inter-barrier regions
  plus a double reflection, $R^{^{2}}\times R^{^{2}}$, in both regions.
 Using $R=-i\,|R|\,e^{i\phi}$, Eq.(\ref{t3}) can be
 rewritten as follows
\begin{equation}
\label{t31}  T^{^{3}}\mbox{\Large $/$}
 \left(1+|R|^{^{\2}}
 e^{^{2i\alpha}}\right)^{\2}=T^{^{3}}\mbox{\Large
 $/$}D^{^{2}}\,\,.
\end{equation}
 However, this
 is by no means all, we can also have reflection from the third barrier,
 backward tunnelling through the second barrier and reflection
 from the first barrier and again tunnelling through the second
 barrier before exiting the structure. This contribution (to
 leading order) yields
$T^{^{5}}\,R^{^{2}}\,e^{^{2ik(L+2d)}}$.
 When all additional internal reflections are allowed for, we
 obtain the partial sum of
\begin{equation}
T^{^{5}}\,R^{^{2}}\,e^{^{2ik(L+2d)}}\left\{1+4\, R^{^{2}}
 e^{^{2ikd}}+ 10\, R^{^{4}}
 e^{^{4ikd}} + ...\right\}\,\,,
 \end{equation}
which leads to
\begin{equation}
\label{t32}
 T^{^{5}}\,R^{^{2}}\,e^{^{2ik(L+2d)}} \mbox{\Large $/$} D^{^{4}} =
 -\,
 T^{^{3}}\,|R|^{^{2}} |T|^{^{2}}\,e^{^{4i\alpha}} \mbox{\Large $/$} D^{^{4}}\,\,.
\end{equation}
The higher denominator power ($D^{4}$) follows from the fact that
each inter-barrier region is crossed twice in the leading
$T^{^{5}}\,R^{^{2}}\,e^{^{2ik(L+2d)}}$ path. This process
continues for the $T^{^{7}}\,R^{^{4}}\,e^{^{4ik(L+2d)}}$ term,
\begin{equation}
\label{t33}
 T^{^{7}}\,R^{^{4}}\,e^{^{4ik(L+2d)}} \, \mbox{\Large
 $/$}D^{^{6}}=T^{^{3}}\,|R|^{^{4}} |T|^{^{4}}\,e^{^{8i\alpha}}
 \mbox{\Large $/$} D^{^{6}}\,\,,
\end{equation}
and so forth. The final result summing Eq.(\ref{t31}),
Eq.(\ref{t32}) and   Eq.(\ref{t33}) is
\begin{equation}
T_{\S}^{^{(3)}}  =  \frac{T^{^{3}}}{D^{^{2}}}\,\left\{1
-\,\frac{|R|^{^{2}} |T|^{^{2}}}{D^{^{2}}}\,e^{^{4i\alpha}} +
\,\frac{|R|^{^{4}} |T|^{^{4}}}{D^{^{4}}}\,e^{^{8i\alpha}} - ...
\right\}
 =   T^{^{3}} \mbox{\Large $/$} \left[D^{^{2}} \left(1+
              \frac{|R|^{^{2}}|T|^{^{2}}}{D^{^{2}}}\,e^{^{4i\alpha}}\right)\right]\,\,.
\end{equation}
To determinate the resonances, in the wave limit, we first take
the modulus squared of the denominator
\[ \left|D^{^{2}}+\, |R|^{^{2}}|T|^{^{2}}e^{^{4i\alpha}}\right|^{^{\,2}}=
1+5\,|R|^{^{4}}+4\,|R|^{^{2}}\left(1+|R|^{^{2}}\right)\cos
(2\alpha)+ 2\,|R|^{^{2}} \cos (4\alpha)\]
 and then differentiate with
respect to $\alpha$, equating to zero,
\[ \sin (2\alpha) \left[1+|R|^{^{2}}+2\,\cos(2\alpha)\right]=0\, \,.\]
This results in\\

\noindent $\bullet$ minimum values of the transmission probability
when  $\sin(2\alpha)=0$, which yields
\begin{eqnarray}
 T_{\S}^{^{(3)}}\left[\cos\alpha=0\right] &=&
T^{^{3}} \mbox{\Large  $/$}\, |T|^{^{2}}\,\,,\nonumber \\
\mathcal{T}_{\mbox{\tiny wave}}^{^{{(3)}}}\left[\cos\alpha=0\right] & = &
|T|^{^{2}}\,\, \\
 T_{\S}^{^{(3)}}\left[\cos\alpha=\pm\,1 \right]
&=& T^{^{3}} \mbox{\Large $/$}\,\left(1+3\,
|R|^{^{2}}\right)\,\,, \nonumber \\
\mathcal{T}_{\mbox{\tiny wave}}^{^{{(3)}}}\left[\cos\alpha=\pm\,1\right] &
= & |T|^{^{2}} \mbox{\Large $/$}\,\left(1+4\,
|R|^{^{2}}/\,|T|^{^{2}}\right)^{\2}\,\,,
\end{eqnarray}

\noindent $\bullet$ maximum values of the transmission probability
when $\cos(2\alpha)=-\left(1+|R|^{^{\2}}\right)/\,2$, which yields
the resonances
\begin{eqnarray}
\mathcal{T}_{\mbox{\tiny
wave}}^{^{(3)}}\left[\cos\alpha=\pm\,|T|/\,2\right]  & = &1
\,\,.
\end{eqnarray}

\section*{\normalsize V. CONTINUITY EQUATIONS AND MATRIX METHOD}

The procedure for calculating the particle terms for the case of
more than three identical barriers becomes very difficult. Not
only must we be careful of coherence but also calculate enough
terms to extrapolate the total series. So, in this section we
provide an alternative procedure based on the matrix method\cite{BORN,Cohen}.
 We derive the expression for the
{\em wave} transmission and reflection amplitudes for $N$
identical barriers, Fig. 1-(d), and then, based upon our results
in the previous sections, we devise a procedure for deriving the
{\em particle} sums. We shall then apply this method to the $N=4$
transmission amplitude.

We first define two $2\times 2$ matrices: $W[\delta,x]$ and the
diagonal $\Delta[\delta x]$,
\[ W[\delta,x]=\left( \begin{array}{rr} e^{\delta x} & e^{-\delta
x}\\ \delta e^{\delta x} & \,\,\,-\delta e^{-\delta x}
\end{array}\right) =W[\delta,0]\,\left( \begin{array}{rr} e^{\delta x} &
0 \\ 0 & \,\,\, e^{-\delta x}
\end{array}\right)=W[\delta,0]\,\Delta[\delta x]\,\,.\]
Let the plane wave solutions in the free region to the left of the
$s$-barrier be
\[ A_s\, e^{ikx}+A'_s\, e^{-\,ikx}\,\,,\]
and that within the barrier
\[ B_s\, e^{\rho x}+B'_s\, e^{-\,\rho x}\,\,.\]
Thus, the plane waves in the extreme left region will be given in
terms of $A_{\0}$ (incoming) and $A'_{\0}$ (reflected). The final
factors on the extreme right of an $N$-barrier system will be
$A_{\N}$ (transmitted) and $A'_{\N}$ (to be set to zero below).

The continuity equations at the two edges of the $s$-barrier read
\begin{eqnarray*}
W[ik,(s-1)(L+d)]\,\left[A_{s-\1}\,\,\,A_{s-\1}'\right]^t & = &
W[\rho,(s-1)(L+d)]\,\left[B_{s}\,\,\,B_{s}'\right]^t\,\,,\\
W[\rho,sL+(s-1)d]\,\left[B_{s}\,\,\,B_{s}'\right]^t & = &
W[ik,sL+(s-1)d]\,\left[A_{s}\,\,\,A_{s}'\right]^t\,\,,\,\,\,\,\,
s=1,2,...N\,\,.
\end{eqnarray*}
Applying successively these matrix equations, we can express the
matrix equation between $(A_{\0}\,,\,\,A'_{\0})$ and
$(A_{\N}\,,\,\,A'_{\N})$,
\begin{eqnarray*}
\left[\begin{array}{l} A_{\0}\\ A_{\0}' \end{array} \right] & =
&\prod_{s=\1}^{\N}
 W^{^{-\1}}[ik,(s-1)(L+d)]\,
W[\rho,(s-1)(L+d)]\,W^{^{-\1}}[\rho,sL+(s-1)d]\,W[ik,sL+(s-1)d]\,\left[\begin{array}{l}
A_{\N}\\ A_{\N}'
\end{array}\right]\\
%& = & \prod_{s=\1}^{\N} & W^{^{-\1}}[ik,(s-1)(L+d)]\,
%W[\rho,0]\,\Delta[-\rho L]\,W^{^{-\1}}[\rho,0]\,W[ik,sL+(s-1)d]\,\,,\\
 & = & \Delta[ikd]\,\left\{\Delta[-ikd] \,W^{^{-\1}}[ik,0]\,
W[\rho,0]\,\Delta[-\rho
L]\,W^{^{-\1}}[\rho,0]\,W[ik,0]\right\}^{\N}
\Delta[ikNL+ik(N-1)d]\,\left[\begin{array}{l} A_{\N}\\ A_{\N}'
\end{array}\right]\,\,.
\end{eqnarray*}
Introducing the matrix $M$,
\[ M = \Delta[-ikd] \,W^{^{-\1}}[ik,0]\,
W[\rho,0]\,\Delta[-\rho L]\,W^{^{-\1}}[\rho,0]\,W[ik,0]\,\,, \]
the previous matrix equation can be rewritten as follows
\begin{equation}
\label{mateq} \left[\begin{array}{l} A_{\0}\\ A_{\0}' \end{array}
\right] = \Delta[ikd]\,M^{^{\N}}\,
\Delta[ikNL+ik(N-1)d]\,\,\left[\begin{array}{l} A_{\N}\\
A_{\N}'\end{array}\right]\,\,.
\end{equation}
A straightforward calculation shows that
\[ M = \left( \begin{array}{rr} F & \,\,\,G^*\\
G & F^* \end{array}\right)\]
with
\[ F =\frac{1}{T\,e^{^{ik(L+d)}}}=\frac{1}{|T|\,e^{^{i\alpha}}}\,\,\,\,\,\mbox{and}\,\,\,\,\,G =
\frac{R\,e^{^{ikd}}}{T\,e^{^{ikL}}}=-\,i\,\frac{|R|}{|T|}\,e^{^{ikd}}
\,\,.\] Observe that $M$ has unit determinant,
$|F|^{^{2}}-|G|^{^{2}}=1$. For an incoming wave from the left,
$A_{\N}'=0$, Eq.(\ref{mateq}) gives
\begin{eqnarray}
A_{\0} & = &
\left(M^{^{\N}}\right)_{\1\1}\,e^{^{ikN(L+d)}}\,A_{\N}\,\, ,
\nonumber \\
A_{\0}' & = &
\left(M^{^{\N}}\right)_{\2\1}\,e^{^{ik[NL+(N-\,2)d\,]}}\,A_{\N}\,\,\,\,,
\end{eqnarray}
where the sub-indices specify the matrix element. Thus, the
reflection and transmission amplitudes are given by
\begin{eqnarray}
R_{\S}^{^{\Np}} & = &
e^{^{-\,2ikd}}\,\left(M^{^{\N}}\right)_{\2\1}\,\mbox{\Large $/$}\,
\left(M^{^{\N}}\right)_{\1\1}\,\,, \nonumber \\
 T_{\S}^{^{\Np}} & = &
e^{^{-\,ikN(L+d)}}\,\mbox{\Large $/$}\,
\left(M^{^{\N}}\right)_{\1\1}\,\, . \label{tsn}
\end{eqnarray}
 For two and three barriers, we have respectively
\[
\left(M^{^{2}}\right)_{\1\1} =
F^{^{2}}+|G|^{^{2}}\,\,\,\,\,\,\,\,\mbox{and}\,\,\,\,\,\,\,\,
\left(M^{^{3}}\right)_{\1\1} =
F\left(F^{^{2}}+|G|^{^{2}}\right)+|G|^{^{2}}\left(F+F^*\right)\,\,.
\]
Consequently, in agreement with our previous
results,
\[
T_{\S}^{^{(2)}}  =   1 \mbox{\Large $/$} \left[
              F^{^{2}}e^{^{2ik(L+d)}}\left(1+\frac{|G|^{^{2}}}{F^{^{2}}}\right)\right] =
               T^{^{2}}\mbox{\Large $/$}\,D\]
and
\[
T_{\S}^{^{(3)}} = 1  \mbox{\Large $/$} \left\{
              F^{^{3}}e^{^{3ik(L+d)}}\left[\left(1+\frac{|G|^{^{2}}}{F^{^{2}}}\right)^{\2}
              +\frac{|G|^{^{2}}}{F^{^{4}}}\right]\right\}
              =   T^{^{3}} \mbox{\Large $/$} \left[D^{^{2}} \left(1+
              \frac{|R|^{^{2}}|T|^{^{2}}}{D^{^{2}}}\,e^{^{4i\alpha}}\right)\right]\,\,
              .
\]
These results are easy to find since we already knew the particle
expressions. We now derive the $N=4$ transmission amplitude. By
using
\[ \left(M^{^{4}}\right)_{\1\1}  =
\left(F^{^{\2}}+|G|^{^{\2}}\right)^{\2}+|G|^{^{\2}}\left(F+F^*\right)^{^{\2}}\,\,,
\]
from Eq.(\ref{tsn}), we get
\begin{equation}
\label{ts4}
 T_{\S}^{^{\d}}  =  1 \mbox{\Large $/$} \left\{
              F^{^{\4}}e^{4ik(L+d)}\left[D^{^{\2}} +
              \frac{|G|^{^{\2}}}{F^{^{\2}}}\left( 1
              +\frac{|F|^{^{\2}}}{F^{^{\2}}}\right)^{^{\2}}\right]\right\}\,\,.
              \end{equation}
The leading term will be $T^{^{\4}}/\,D^{^{\3}}$. {\em This term
must be factorized}. The remaining expression can and must be
expressed in terms of $|R|^{^{2}}$, $|T|^{^{2}}$ and $e^{^{2 i
\alpha}}$. By adding and subtracting, in the square bracket of the
denominator of Eq.(\ref{ts4}) a $D^{^{3}}$  term,  and by
recalling that $|F|^{^{2}}=1+|G|^{^{2}}$, we can rewrite the
transmission amplitude as follows
\begin{eqnarray}
T_{\S}^{^{(4)}}
              & = &1 \mbox{\Large $/$} \left\{
              F^{^{4}}e^{^{4ik(L+d)}}\left[D^{^{3}}
              -\frac{|G|^{^{2}}}{F^{^{2}}}D^{^{2}}+
              \frac{|G|^{^{2}}}{F^{^{2}}}\left( D
              +\frac{1}{F^{^{2}}}\right)^{\2}\right]\right\} \nonumber \\
              & = &1 \mbox{\Large $/$} \left[
              F^{^{4}}e^{^{4ik(L+d)}}\left(D^{^{3}} + 2\,
              \frac{|G|^{^{2}}}{F^{^{4}}}D+
              \frac{|G|^{^{2}}}{F^{^{6}}} \right)\right] \nonumber \\
              & = &T^{^{4}} \mbox{\Large $/$}\left[D^{^{3}} \left(1 + 2\,
              \frac{|R|^{^{2}} |T|^{^{2}}}{D^{^{2}}}\,e^{^{4i\alpha}} +
              \frac{|R|^{^{2}} |T|^{^{4}}}{D^{^{3}}}\,e^{^{6i\alpha}}
              \right)\right]\,\, .
\end{eqnarray}
To obtain the particle sum, we must explicit the denominator as a
series in the numerator. For $N=4$, we find up order $|T|^{^{8}}$,
\begin{equation}
\frac{T^{^{4}}}{D^{^{3}}}\, \left\{1 -
2\,|R|^{^{2}}\,e^{^{4i\alpha}}\,
              \frac{ |T|^{^{2}}}{D^{^{2}}} +
              \left(3\,|R|^{^{4}}\,e^{^{8i\alpha}}-|R|^{^{2}}\,e^{^{6i\alpha}}\right)\,
              \frac{|T|^{^{4}}}{D^{^{4}}}
              +\, ... \right\}\,\, .
\end{equation}
For the individual particle contributions we must expand the $D$
terms in the numerator. The above result agrees with a direct, but
more tedious, particle calculation to this order. The maximum and
minimum values of the transmission probability are obtained by
following the procedure of the
previous section. This results in\\

\noindent $\bullet$ minimum values of the transmission probability

\begin{eqnarray}
\mathcal{T}_{\mbox{\tiny
wave}}^{^{{\d}}}\left[\cos\alpha=\pm\,|T|/\sqrt{6}\right] & = &
|T|^{^{\2}} \mbox{\Large $/$}\,\left(1+5\,
|R|^{^{\2}}/\,27\right)\,\,, \\
\mathcal{T}_{\mbox{\tiny wave}}^{^{{\d}}}\left[\cos\alpha=\pm\,1\right] &
= & 1\, \mbox{\Large $/$}\,\left(1+8\,
|R|^{^{\2}}/\,|T|^{^{\4}}\right)^{\2}\,\,,
\end{eqnarray}

\noindent $\bullet$ maximum values of the transmission probability

\begin{equation}
\mathcal{T}_{\mbox{\tiny wave}}^{^{{\d}}}\left[\cos\alpha=0\right] =
\mathcal{T}_{\mbox{\tiny
wave}}^{^{{\d}}}\left[\cos\alpha=\pm\,|T|/\sqrt{2}\right]  =
1\,\,.
\end{equation}

\section*{\normalsize VI. CONCLUSIONS}

We have considered in this paper tunnelling through two, three and
four identical barriers. Our approach is that of considering the
reflection and transmission amplitudes successively for each
barrier, the so called {\em particle approach}. This is directly
relevant to situations in which the incoming (single) wave packet
is small compared to the inter-barrier distance $d$. However, it
also yields the wave limit if all particle contributions are
summed and considered as a single outgoing reflected and
transmitted wave. In this wave limit one encounters resonance
phenomena. We have re-derived the condition for resonances and
expressed them in terms of $\cos \alpha$. Resonances do not occur
in the particle limit. We have exhibited numerically the
transition from the wave to particle limit for the case of twin
tunnelling by plotting the transmission probability as a function
of the inter-barrier distance $d$. In the particle limit
oscillations die out.

The calculation of the individual particle terms becomes
cumbersome for four barriers and higher. Care must be taken to sum
coherent contributions which are produced from different paths
within the potential structure. For these cases it is simpler to
invert our procedure and first calculate the wave limit amplitude.
This has been done, as an example, for the four barrier case in
the previous section with the help of a matrix method and the
rules given for its representation in ``particle''  form, i.e. in
terms of $R$ and $T$ and powers of $e^{^{2 i\alpha}}$. A series
expansion of the denominators completes the procedure.\\

There are a number of consequence of our analysis which merit a
mention:\\

\noindent 1) The resonance conditions for multiple identical
barrier tunnelling will automatically extrapolate into the case of
$E>V_{\0}$, i.e. for above barrier diffusion. Thus, there are two
classes of above barrier resonances.\\

1-a) When above barrier resonance occurs for a single barrier,
$|T|=1$, independently of the value of $d$ we have
$T_{\S}^{^{\b}}=T^{^{\2}}$, where $T$ is now the above barrier
expression, i.e. $\rho \to i\, q$ with
$q=\sqrt{2m(E-V_{\0})}/\hbar$ in Eq.(\ref{teq}).\\

1-b) When $|T|\neq 1$ but $\phi+kd=(n+1/2)\,\pi$ as in the
tunnelling case. These are additional resonances corresponding to
constructive interference effects of the twin barriers possible in
the wave limit even if each single barrier is not resonant.\\

\noindent 2) For twin barriers, the conditions for tunnel
resonance requires {\em identical} barriers. For two different
barriers, either in width or height or both, the formula for the
transmission amplitude  becomes
\[ T_{\S}^{^{\bd}} = T_{\1}T_{\2}\mbox{\Large $/$} \left(
1-R_{\1}R_{\2}\,e^{2ikd}\right) \,\,, \] with the indices
indicating the individual barrier amplitudes. Now, for no value of
$kd$ will this be of unit modulus. We expect the same to
happen also for higher numbers of barriers.\\

\noindent 3) Consider twin tunnelling at resonance. The phase of
$T_{\S}^{^{\b}}$ is totally given by the phase of $T^{^{\2}}$. The
time calculated at $x=2L+d$, the exit point for transmission, is
\[t=\left[
\frac{\partial k}{\partial E}\,\left(x-2L\right) + 2\,
\frac{\partial \phi}{\partial E}\right]_{x=2L+d,\,\,k=k_{\0}}=
\frac{d}{v_g} + 2\, \left[\frac{\partial \phi}{\partial
E}\right]_{k=k_{\0}}\,\,,
\]
where $v_g$ is the group velocity in the free space. Thus, due to
the linear term in $d$, there is no genera\-lized Hartman
effect\cite{DB} at resonance.\\

\noindent Finally, we wish to place our results in a broader
context. The existence of wave and particle limits is related to
the relative sizes of the incoming wave packet and the size of the
potential structure. The results of this paper are an example of
the physical relevance of wave packet dimensions. Interesting applications
occurs in particle oscillation phenomena\cite{D1,D2,D3,D4},
relativistic tunneling\cite{D5,D6,D7}, laser interaction with dielectric blocks\cite{D8,D9},
and, generally speaking, in all interference based results. These
features are generally neglected in the literature where only
single plane waves are used. Plane waves are a legitimate
operational tool, but working {\em only} with single plane waves
(infinite wave packet size) may obfuscate significant physical
insight. \\

\section*{\small \rm ACKNOWLEDGEMENTS}
One of the authors (SdL) wish to thank the
Department of Physics, University of Salento (Lecce, Italy), where the paper
was written, for the invitation and the hospitality. He also thanks the FAPESP
(Brazil) for financial support by the grant n. 10/02216-2. The authors also thank
the anonymous referees for their constructive comments  and useful suggestions.

\newpage

\begin{figure}[hbp]
\hspace*{-2.5cm}
\includegraphics[width=19cm, height=22cm, angle=0]{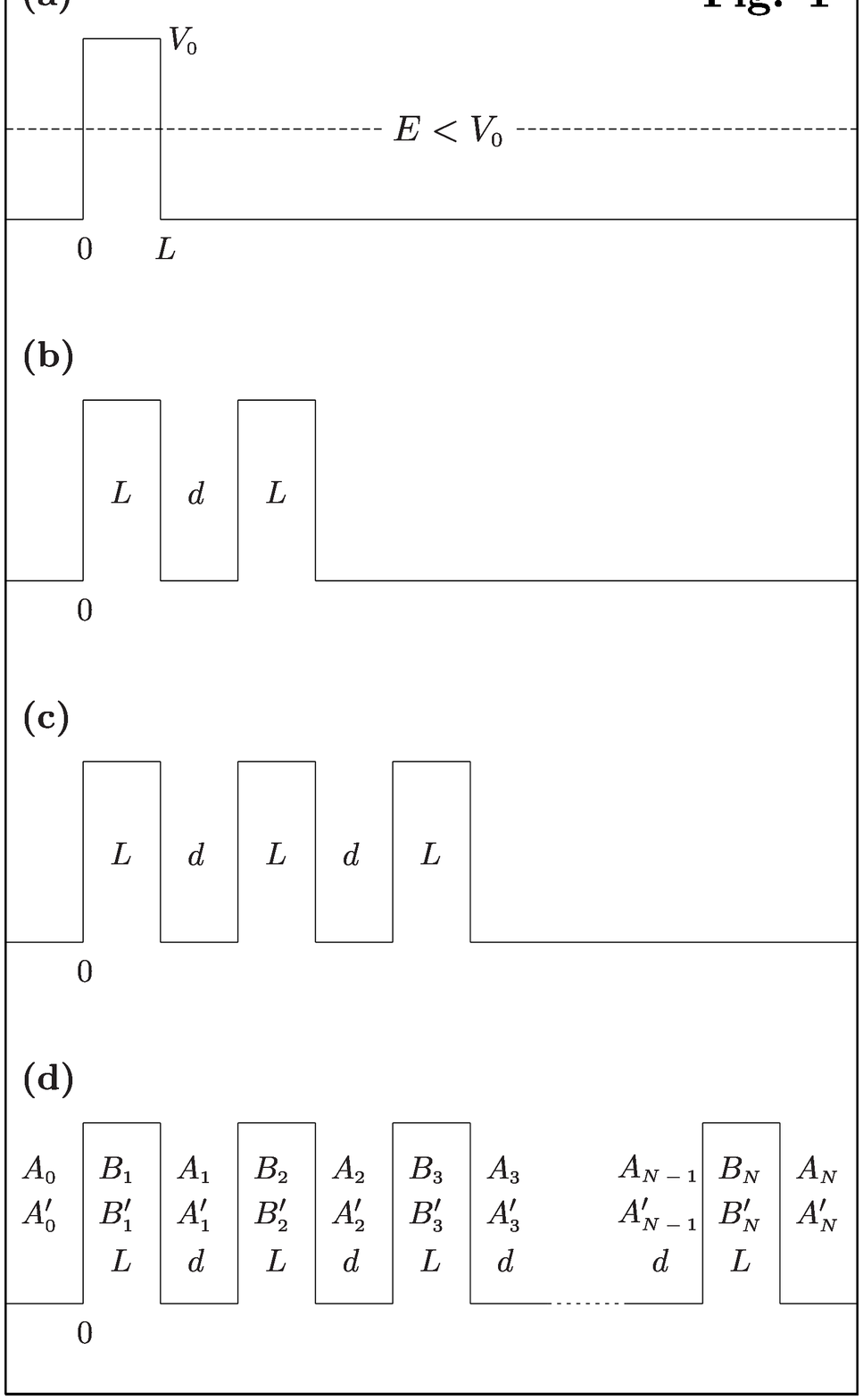}
\vspace*{-2cm}
 \caption{Potential shapes: (a) single barrier, (b) twin barriers,
 (c) triple identical barrier and (d) general structure for matrix
 calculation. In (d)
 the $A/A'$ and $B/B'$ terms indicate the amplitudes in the free and potential regions.}
\end{figure}

\newpage

\begin{figure}[hbp]
\hspace*{-2.5cm}
\includegraphics[width=19cm, height=22cm, angle=0]{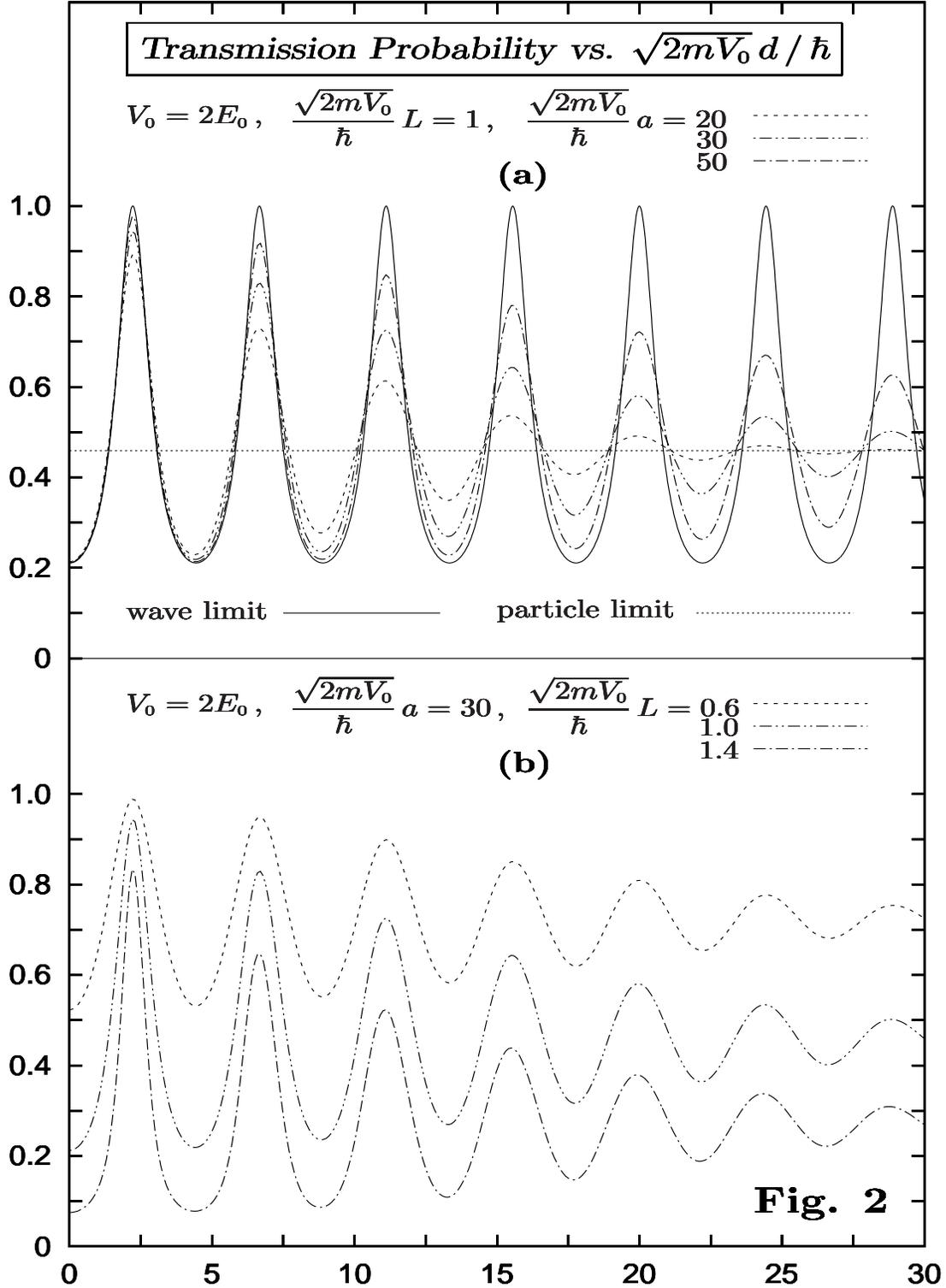}
\vspace*{-2cm}
 \caption{Transmission probability vs the inter-barrier distance.
 Upper curves are for a fixed barrier width ($L$) and various wave packet
 widths ($a$).
 Lower curves are for a fixed wave packet width and various barrier
 widths.   The curves show the gradual transition from the wave limit (continuous line) to the particle limit (dotted line) exhibited by the damping of the oscillations. Observe that the particle limit is the same for all the curves plotted in Fig.2-a whereas each particle limit is different in Fig.2-b. This is due to the fact that $\mathcal{T}_{\mbox{\tiny particle}}^{^{{\b}}}$ depends on the barrier width.
 }
\end{figure}

\newpage

\begin{figure}[hbp]
\hspace*{-2.5cm}
\includegraphics[width=19cm, height=22cm, angle=0]{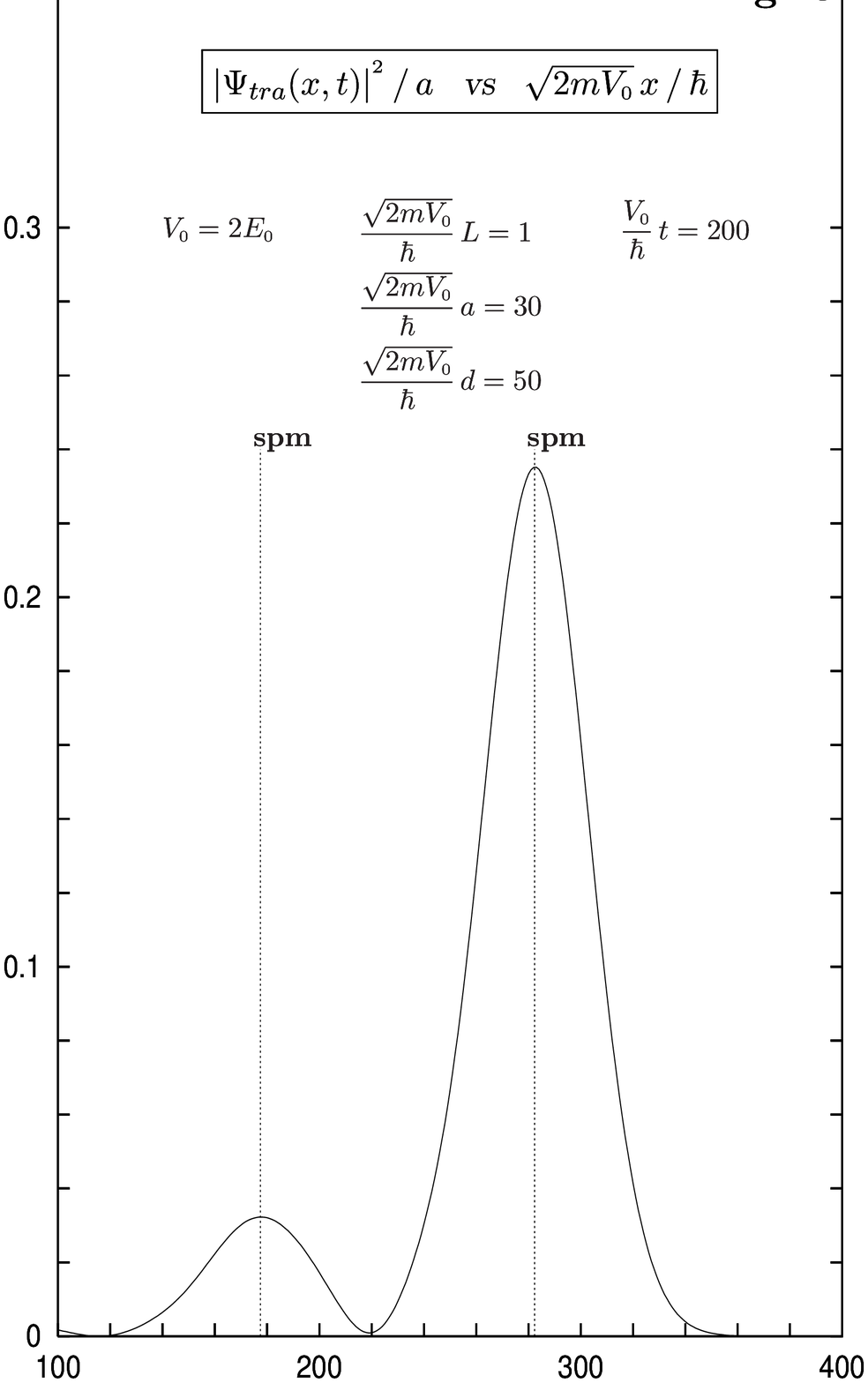}
\vspace*{-2cm}
 \caption{The density probability for the transmitted wave is plotted as a function of $x$ for a
 fixed time. This time is such that two outgoing waves are seen. The SPM estimates (vertical lines)
are in excellent agreement with the numerical calculation.}
\end{figure}

\end{document}